\documentclass[prl, twocolumn, superscriptaddress, showpacs,amsmath,amssymb]{revtex4}

\usepackage{float}
\usepackage{graphicx}
\usepackage{subfigure}
\usepackage{amssymb}
\usepackage{amsfonts}

\begin{document}

\title{Photon Antibunching from a Single Quantum Dot-Microcavity System in the Strong Coupling Regime}
\author{David Press}
\email{dlpress@stanford.edu}
\author{Stephan G\"{o}tzinger}
\affiliation{Edward L.
    Ginzton Laboratory, Stanford University, Stanford, California
    94305-4085, USA}
\author{Stephan Reitzenstein}
\author{Carolin Hofmann}
\author{Andreas L\"{o}ffler}
\author{Martin Kamp}
\author{Alfred Forchel}
\affiliation{Technische Physik, Universit\"{a}t W\"{u}rzburg, Am
Hubland, D-97074 W\"{u}rzburg, Germany}
\author{Yoshihisa Yamamoto}
\altaffiliation[Also at ]{National Institute of Informatics, Tokyo,
Japan} \affiliation{Edward L.
    Ginzton Laboratory, Stanford University, Stanford, California
    94305-4085, USA}

\begin{abstract}

We observe antibunching in the photons emitted from a
strongly-coupled single quantum dot and pillar microcavity in
resonance.  When the quantum dot was spectrally detuned from the
cavity mode, the cavity emission remained antibunched, and also
anticorrelated from the quantum dot emission. Resonant pumping of
the selected quantum dot via an excited state enabled these
observations by eliminating the background emitters that are usually
coupled to the cavity. This device demonstrates an on-demand single
photon source operating in the strong coupling regime, with a
Purcell factor of $61\pm7$ and quantum efficiency of 97\%.

\end{abstract}

\maketitle

Cavity quantum electrodynamics (CQED), addressing the interaction
between a quantum emitter and a cavity, has been a central topic in
atomic physics for
decades~\cite{Kleppner81a,Haroche83as,Haroche85a,Kimble92a} and has
recently come to the forefront of semiconductor
physics~\cite{Forchel04a,Gibbs04a,Gerard05a,Gibbs06a}.  If the
coupling between the single quantum emitter and cavity mode is
strong compared to their decay rates, the emitter and cavity
coherently exchange energy back and forth leading to Rabi
oscillations. This strong coupling (SC) regime is of great interest
for a variety of quantum information applications, especially with a
solid-state implementation. A SC QD-microcavity system could lead to
a nearly ideal single photon source (SPS) for quantum information
processing, with extremely high efficiency and photon
indistinguishability~\cite{Imamoglu04a}. The same technology could
be applied as an interface between a spin qubit and single photon
qubit in a quantum network~\cite{Sham05a}.

SC between a single atom and a cavity was first achieved more than a
decade ago~\cite{Kimble92a}. An analogous system in the solid-state
is the excitonic transition of a semiconductor quantum dot (QD)
together with a semiconductor microcavity. Several groups have
recently reported SC between a single (In,Ga)As QD and either
micropillar~\cite{Forchel04a}, photonic crystal~\cite{Gibbs04a}, or
microdisk~\cite{Gerard05a} resonators. SC can also occur between a
single cavity mode and a collection of degenerate emitters, such as
an ensemble of atoms or a quantum well~\cite{Arakawa92as}. However,
in the latter case the behavior is classical: adding or removing one
emitter or one photon from the system has little effect.

In previous studies of QD-cavity
SC~\cite{Forchel04a,Gibbs04a,Gerard05a} it was argued that the
spectral density of QDs was sufficiently low that it is unlikely
that several degenerate emitters contributed to the anticrossing.
However, it was not verified that the system had one and only one
emitter. There was a surprisingly large amount of emission from the
cavity mode when the QD was far detuned. It was unclear whether this
emission originated from the particular single QD or from many
background emitters.  An important step to establish SC in
solid-state CQED is verification that the double-peaked spectrum
originates from a single quantum emitter, not a collection of
emitters, interacting with the cavity mode.

In this Letter we present proof that the emission from a
strongly-coupled QD-microcavity system is dominated by a single
quantum emitter. Photons emitted from the coupled QD-microcavity
system at resonance showed a high degree of antibunching.  Away from
resonance, emission from the QD and cavity modes was anticorrelated,
and the individual emission lines were antibunched. The key to these
observations was to resonantly pump the selected QD via an excited
QD state to prevent background emitters from being excited. These
background emitters, which are usually excited by an above-band
pump, prevent the observation of antibunching by emitting photons
directly into the cavity mode and by repeatedly exciting the QD
after a single laser pulse. With pulsed resonant excitation, the
device demonstrates the first solid-state single photon source
operating in the strong coupling regime. The Purcell factor exceeds
60 and implies very high quantum efficiency, making such a device
interesting for quantum information applications.

Planar cavities were grown with Bragg mirrors consisting of 26 and
30 pairs of AlAs/GaAs layers above and below a GaAs cavity. A layer
of InGaAs QDs, with an indium content of about 40\% and a density of
$10^{10}$~cm$^{-2}$, was grown in the central antinode of the
cavity. The QDs typically show splittings between the \emph{s-}shell
and \emph{p-}shell transition energies of $25-30$~meV, suggesting
lateral QD dimensions of $20-30$~nm~\cite{Maksym01as}. The cavities
were etched into circular micropillars with diameters varying from 1
to 4$~\mu$m. An electron microscope image of a 1.2$~\mu$m diameter
micropillar is shown in Fig.~1(a). Further details on fabrication
can be found in Ref.~\cite{Forchel05a}.

The coupled-oscillator model gives the complex eigen-energies of the
system's two normal modes:
\begin{equation}
     E_{1,2} = \frac{E_c+E_x}{2}-i\frac{\gamma_c+\gamma_x}{4} \pm \sqrt{ g^2-\frac{(\gamma_c-\gamma_x-2i\Delta)^2}{16}}
\end{equation}

where $E_x$ and $E_c$ are the energies of the QD exciton
and cavity modes, $\gamma_{x}$ and $\gamma_{c}$ are their full-width
half-maxima (FWHM), $\Delta=E_x-E_c$ is the detuning, and $g$ is the
exciton-cavity coupling strength. SC requires
$g^{2}>(\gamma_{c}-\gamma_{x})^{2}/16$, which leads to a splitting
of the two eigen-energies at resonance ($\Delta=0$) by an amount
called the vacuum Rabi splitting. For typical QDs and semiconductor
microcavities, $\gamma_{x}$ (a few $\mu$eV) is much smaller than
$\gamma_{c}$ ($\sim100~\mu$eV), and the SC condition reduces to
$g>\gamma_c/4$.  In order to reach SC with a given oscillator
strength one must maximize the ratio of cavity quality factor to
mode volume, $Q/\sqrt{V}$~\cite{Forchel04a,Gibbs04a}. Our sample
showed the highest $Q/\sqrt{V}$ ratio for $1.8~\mu$m diameter
pillars, which typically exhibited $Q\sim10000-20000$, with a mode
volume $V_{m}\sim0.43~\mu$m$^{3}$.

Photoluminescence (PL) measurements were performed while the sample
was cooled to cryogenic temperatures. Increasing the sample
temperature caused the QD excitons to red-shift faster than the
cavity mode, allowing the QDs to be tuned by nearly 1.5~nm relative
to the cavity between 6~K and 40~K. The sample was optically pumped
by a tunable continuous wave (CW) or mode-locked pulse Ti:sapphire
laser, focused to a 2~$\mu$m spot through a 0.75 $NA$ objective. PL
was detected by a 750mm grating spectrometer with N$_{2}$-cooled CCD
(spectral resolution 0.03~nm). For photon correlation measurements
the PL was spectrally filtered by a 0.2~nm resolution monochromator
before entering a Hanbury-Brown-Twiss setup~\cite{Santori01as}.
Lifetime measurements were performed using a streak camera with
temporal resolution of 25~ps.

In the simplest picture, above-band pumping creates electron-hole
pairs that can radiatively recombine to emit photons at the QDs'
quantized energy levels. The cavity should be nearly dark if no QD
level is resonant with it. However, in previous studies of QD SC the
cavity emission was much brighter than the QD emission even when no
QD was resonant with the
cavity~\cite{Forchel04a,Gibbs04a,Gerard05a}. It was unclear whether
the cavity emission resulted from coupling to the specific QD
involved in SC, or to a broad background of emitters such as
spectrally far-detuned QDs and wetting layer states. These
background emitters might contribute to the cavity emission by
simultaneously emitting a cavity photon and one or more phonons.

In order to eliminate any background emitters, the laser can be
tuned to resonantly pump the excited state (\emph{p}-shell) exciton
in a selected QD~\cite{Santori01as}. The exciton quickly thermalizes
to the QD ground state (\emph{s}-shell) where it can interact with
the cavity. Ideally, resonant pumping creates excitons only in the
selected QD, eliminating all extraneous emitters coupled to the
cavity.

\begin{figure}[t]
\centerline{\includegraphics[width=9cm]{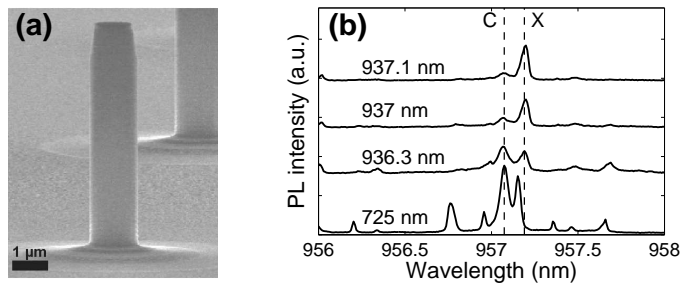}} \caption{(a)
SEM image of a 1.2 $\mu$m diameter pillar cavity. (b) Above-band
pumping compared to resonant pumping of a chosen QD in Pillar 1.
With above-band pump (725~nm, 0.4~$\mu$W), the chosen QD exciton (X)
emits, but so do the cavity (C) and many other QDs. With 937.1~nm
(3~$\mu$W) pump, the chosen QD is selectively excited and its PL
dominates an otherwise nearly flat spectrum.}
\end{figure}

The PL spectrum of a typical weak-coupling device called Pillar 1,
excited by CW above-band pumping, is shown in the lowest trace in
Fig.~1(b). The cavity mode ($Q=17300$) could be identified amongst
the various QD lines by its broader linewidth, slower tuning with
respect to temperature, and lack of saturation at high pump powers.
The cavity emits strongly even though there is no QD resonant. The
higher traces in Fig.~1(b) show how tuning the pump laser towards an
excited state in a chosen QD (937.1~nm in this case) can selectively
excite the QD with greatly reduced background cavity emission.
Resonant pumping suppresses the cavity emission relative to the QD
emission by roughly a factor of ten in this particular pillar. The
resonant pump was nearly ten times as intense as the above-band pump
to achieve the same PL intensity, which caused local heating and
lead to a slight red shift ($0.01-0.03$~nm) of the QD line.

The temperature dependent PL for a device exhibiting SC called
Pillar 2 is presented if Fig.~2. A clear anticrossing of the QD line
and the cavity mode at resonance is evident. When the device was
pumped above-band (725~nm), the cavity was significantly brighter
than the QD and many QDs lines were visible. Resonant pumping of the
particular QD involved in SC eliminated the other QD lines and
reduced the cavity background emission. The vacuum Rabi splitting at
resonance is more pronounced with resonant pumping, possibly because
the above-band pump creates background excitons and trapped charges
that interact with the QD exciton to broaden its emission.

\begin{figure}[t]
\centerline{\includegraphics[width=9cm]{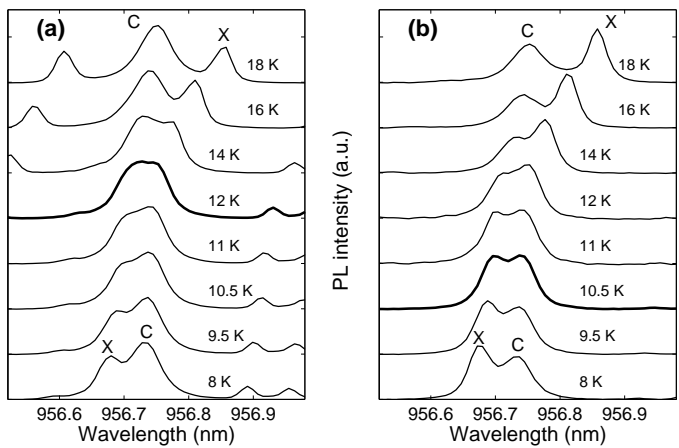}}
\caption{Temperature dependent PL from Pillar 2 with (a) above-band
CW pump (725~nm), and (b) resonant CW pump ($936.25-936.45$~nm).
Each spectrum is rescaled to a constant maximum since tuning the QD
changes excitation efficiency. Resonance occurred at lower
temperature for resonant pump case (10.5~K vs. 12~K) due to local
heating.}
\end{figure}

\begin{figure}[b]
\centerline{\includegraphics[width=9cm]{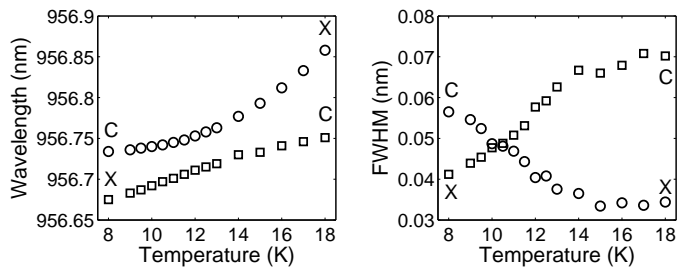}} \caption{
Emission wavelength and FWHM of upper (circles) and lower (squares)
lines as a function of temperature, based on double-Lorentzian fits
to resonantly-excited spectra of Pillar 2 (Fig.~2(b)).}
\end{figure}

The line centers and linewidths of the resonantly-pumped QD-cavity
system (Fig.~2(b)) are shown in Fig.~3. For the lowest temperatures
the lower line is narrower and exciton-like, and the upper line is
broader and cavity-like. Increasing the temperature causes the lines
to switch character as they anticross. From Fig.~3 we determine the
cavity linewidth of Pillar 2 is $\gamma_c=85~\mu$eV ($Q=15200$) and
the vacuum Rabi splitting is 56~$\mu$eV. Using formula (1) we
calculate $g=35~\mu$eV. This gives a ratio of
$g/\gamma_c=0.41>\frac{1}{4}$ as required to satisfy the strong
coupling condition. Fits to the above-band pumped spectra yield a
similar value for $\gamma_c$, and slightly smaller values for the
vacuum Rabi splitting (50~$\mu$eV) and $g$ (33~$\mu$eV).

To verify the quantum nature of the system and determine whether a
single emitter is responsible for the photon emission, we measured
the photon autocorrelation function $g^{(2)}(\tau)=\langle
I(t)I(t+\tau) \rangle / \langle I(t) \rangle^{2}$ of the PL from
Pillar 2. With weak excitation, the width of the dip in
$g^{(2)}(\tau)$ near $\tau=0$ is given by the lifetime of the
emitter, which is roughly 15~ps (i.e. twice the cavity lifetime) for
the resonantly-coupled QD-cavity system. The emitter's extremely
fast decay rate necessitates a pulsed excitation scheme since
conventional photon counters cannot resolve such a short time scale.

\begin{figure}[b]
\centerline{\includegraphics[width=9cm]{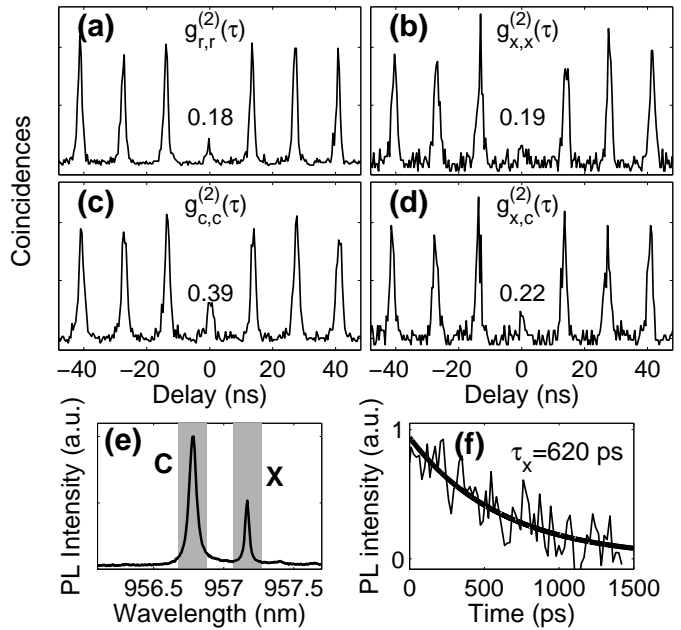}} \caption{(a)
Measured autocorrelation function of the SC system at resonance,
$g^{(2)}_{r,r}(0)=0.18$. (b-e) QD detuned 0.4~nm from cavity. (b)
Autocorrelation function of QD emission only,
$g_{x,x}^{(2)}(0)=0.19$. (c) Autocorrelation function of cavity
emission only, $g_{c,c}^{(2)}(0)=0.39$. (d) Cross-correlation
function of QD and cavity, $g_{x,c}^{(2)}(0)=0.22$. (e) PL spectrum.
Shaded regions indicate pass-bands of spectral filter for
correlation measurements. (f) Lifetime measurement of QD only,
detuned 0.7~nm from cavity. Dark count backgrounds have been
subtracted.}
\end{figure}

The autocorrelation function of photons collected from the coupled
QD-cavity system at resonance is shown in Fig.~4(a). The observed
value of $g^{(2)}_{r,r}(0)=0.18<\frac{1}{2}$ proves that the
emission from the coupled QD-cavity is dominated by the single QD
emitter. Increasing pump power yielded higher values for
$g^{(2)}(0)$ as the QD saturated but the cavity emission continued
to rise.

Next the QD was red detuned by 0.4~nm from the cavity mode so that
photon statistics could be collected from the cavity and QD emission
lines separately. Surprisingly, even with the resonant pump tuned to
selectively excite the chosen QD, the cavity emission was $\sim$3.5
times brighter than the QD (see Fig.~4(e)). (Note that with
above-band pumping, background emitters were excited and the cavity
emission grew another five times brighter relative to the QD).  The
QD emission was antibunched as expected with $g^{(2)}_{x,x}(0)=0.19$
(Fig.~4(b)). Interestingly, the cavity emission was also antibunched
with $g^{(2)}_{c,c}(0)=0.39<\frac{1}{2}$ (Fig.~4(c)), showing that
the cavity emission is dominated by a single quantum emitter. This
slightly higher value of $g^{(2)}(0)$ suggests that some background
emitters were still weakly excited and contribute to the cavity
emission. Finally, the cross-correlation function between the QD
exciton and cavity emission $g^{(2)}_{x,c}(\tau)$ was measured
(Fig.~4(d)). Strong antibunching was observed with
$g^{(2)}_{x,c}(0)=0.22$, conclusively proving that the single QD
emitter is responsible for both peaks in the PL spectrum.

The bright cavity emission cannot be explained by radiative coupling
to the QD due to their large detuning.  This suggests that another,
unidentified mechanism couples QD excitations into the cavity mode
when off-resonance, in agreement with another recent report on
QD-cavity SC~\cite{Imamoglu06a}.  This coupling could possibly be
mediated by the absorption or emission of thermally-populated
acoustic phonons~\cite{Gerard03a,Zimmermann04a}.

When another SC pillar was pumped with above-band pulses,
$g^{(2)}_{r,r}(0)$ of the resonantly-coupled QD-cavity system
remained between 0.85 and 1 even for the lowest pump powers.
Antibunching could not be observed with an above-band pump for two
reasons. First, the above-band pump creates many background emitters
that couple to the cavity mode, as discussed above. Second, the free
excitons created by the pump have lifetimes much longer than the
coupled QD-cavity lifetime, allowing multiple capture and emission
processes after a single laser pulse. Resonant pumping solves both
of these problems.

Under pulsed resonant excitation at the resonance temperature,
Pillar 2 emits a pulse train of photons, demonstrating the first
solid-state SPS operating in the SC regime. A useful figure of merit
for a SPS is the Purcell factor $F_P$. In the weak coupling limit,
$F_P$ gives the enhancement of the QD's emission rate $\gamma$ due
to the cavity: $\gamma=(1+F_P)\gamma_x$. This relation no longer
holds in the SC regime, where the decay rates of the coupled
QD-cavity states are fixed at $(\gamma_c+\gamma_x)/2$. We define the
Purcell factor more generally as $F_P = \frac{4g^2}{\gamma_c
\gamma_x}$ (also called the cooperativity parameter in atomic
physics), where $\gamma_x$ is the QD's emission rate in the limit of
large detuning from the cavity. This Purcell factor is often used to
quantify the performance of CQED-based quantum information
processing schemes~\cite{Imamoglu04a, Fattal06a}, and is related to
the quantum efficiency of the resonantly-coupled
SPS~\cite{Raymer05a}:
\begin{equation}
\eta = \frac{F_P}{1+F_P}\frac{\gamma_c}{\gamma_c+\gamma_x}
\end{equation}
The efficiency $\eta$ gives the probability that a photon will be
emitted into the cavity mode given that the QD is initially excited.
We measured the QD lifetime to be $620\pm70$~ps when the QD was
detuned by 0.7~nm from the cavity mode, as shown in Fig.~4(c).  At
this moderate detuning, the QD's emission rate was slightly enhanced
from $\gamma_x$ by coupling to the cavity. We may calculate the
decay rate $\gamma_x=1/\tau_x$ from formula (1) using
$\gamma_{1,2}(\Delta)=2\mbox{Im}\{E_{1,2}\}$. From this expression
and the measured lifetime, we determine the QD's lifetime in the
large detuning limit to be $\tau_x=700\pm80$~ps. This lifetime
agrees with measurements of bulk QDs showing an ensemble lifetime of
600~ps when we consider that a pillar microcavity may quench the
emission rate of a far-detuned QD by roughly 10\%~\cite{Forchel01a}.
Using $\tau_x$ we determine a Purcell factor of $61\pm7$ and quantum
efficiency of $97.3\pm0.4$\%.

The high quantum efficiency and short single-photon pulse duration
make this device directly applicable to high speed quantum
cryptography. However, the incoherent nature of the resonant pump
likely results in moderate photon indistinguishability of around
50\%. Indistinguishability could be improved using a coherent pump
scheme, such as one involving a cavity-assisted spin flip Raman
transition~\cite{Imamoglu04a,Sham05a,Fattal06a}, to make the device
ideal for quantum information processing with single photons.

In conclusion, we have observed antibunching in the photon
statistics from a strongly coupled QD-microcavity system. The
suppressed value of $g^{(2)}(0)=0.18$ from the system at resonance
proves that a single quantum emitter dominates the photon emission.
Off-resonance, the QD and cavity emission were both antibunched as
well as anti-correlated, further confirming that only one emitter is
responsible for the PL. Resonant pumping was essential to these
observations, since it eliminated the background emitters that can
scatter photons directly into the cavity mode and repeatedly excite
the QD after a single laser pulse. Our results demonstrate a
solid-state single photon source operating in the strong coupling
regime, with a Purcell factor of $61\pm7$ and quantum efficiency of
97\%.

We thank D. Fattal for many helpful discussions.  Financial support
was provided by the MURI Center for Photonic Quantum Information
Systems (ARO/ARDA Program No. DAAD19-03-1-0199), JST/SORST, NTT
Basic Research Laboratories, the Deutsche Forschungsgemeinschaft via
Research Group Quantum Optics in Semiconductor Nanostructures, the
European Commission through the IST Project QPhoton, and the State
of Bavaria. D.~P. thanks Sony Corporation, and S.~G. thanks the
Humboldt Foundation.

\bibliographystyle{apsrev}

\end{document}